\documentclass[conference]{IEEEtran}
\IEEEoverridecommandlockouts
\usepackage{cite}
\usepackage{amsmath,amssymb,amsfonts}
\usepackage{algorithmic}
\usepackage{graphicx}
\usepackage{textcomp}
\usepackage{xcolor}
\usepackage{listings}
\usepackage{url}

\usepackage[colorlinks=true,linkcolor=blue,citecolor=blue]{hyperref}
\usepackage{subfigure}

\begin{document}

\title{Oobleck: Low-Compromise Design for Fault Tolerant Accelerators}

\author{\IEEEauthorblockN{Guy Wilks}
\IEEEauthorblockA{Carnegie Mellon University} 
\and
\IEEEauthorblockN{Brian Li}
\IEEEauthorblockA{UC Santa Barbara} 
\and
\IEEEauthorblockN{Jonathan Balkind}
\IEEEauthorblockA{UC Santa Barbara}}

\maketitle

\begin{abstract}
Data center hardware refresh cycles are lengthening. However, increasing processor complexity is raising the potential for faults. To achieve longevity in the face of increasingly fault-prone datapaths, fault tolerance is needed, especially in on-chip accelerator datapaths. Previously researched methods for adding fault tolerance to accelerator designs require high area, lowering chip utilisation. We propose a novel architecture for accelerator fault tolerance, Oobleck, which leverages modular acceleration to enable fault tolerance without burdensome area requirements.

In order to streamline the development and enforce modular conventions, we introduce the Viscosity language, an actor based approach to hardware-software co-design. Viscosity uses a single description of the accelerator's function and produces both hardware and software descriptions.

Our high-level models of data centers indicate that our approach can decrease the number of failure-induced chip purchases inside data centers while not affecting aggregate throughput, thus reducing data center costs. To show the feasibility of our approach, we show three case-studies: FFT, AES, and DCT accelerators. We additionally profile the performance under the key parameters affecting latency. Under a single fault we can maintain speedups of between $1.7\times-5.16\times$ for accelerated applications over purely software implementations. We show further benefits can be achieved by adding hot-spare FPGAs into the chip.

\end{abstract}

\date{November 15th 2023}

\section{Introduction}

As Moore's Law comes to an end~\cite{doi:10.1126/science.aam9744}, the steady performance improvements which came from transistor miniaturization will slow. Even if improvements are made to fit more transistors on a single integrated circuit, those improvements will come less often and at higher development costs. Naturally, the incentive to upgrade processors as often will dramatically decrease. This trend is already observable in large data center contexts. In 2020, Microsoft, Google, and Meta had three year hardware life-cycles. In 2022 and 2023, they all increased their hardware refresh cycles to either five or six years \cite{Caballar_2022, McCurdy_2022, Horizon_Editorial_2023}.

Longer hardware refresh cycles mean that the hardware itself will have to last longer. However, in data center contexts, various device errors can occur, cutting the life of the processors short~\cite{51477, DBLP:journals/corr/abs-2102-11245, 10.1145/3600006.3613149}. As the number of processors in data centers increases, non-transient faults are statistically more likely to crop up. This effect is exacerbated by wider and more dense data paths that increase the likelihood of faults~\cite{DBLP:journals/corr/abs-2102-11245}. This increase in processor faults can be partially attributed to increasing complexity in architectural design and smaller feature sizes which push the limits of CMOS scaling~\cite{10.1145/3458336.3465297}.
\begin{figure}
    \centering
    \includegraphics[width=0.45\textwidth]{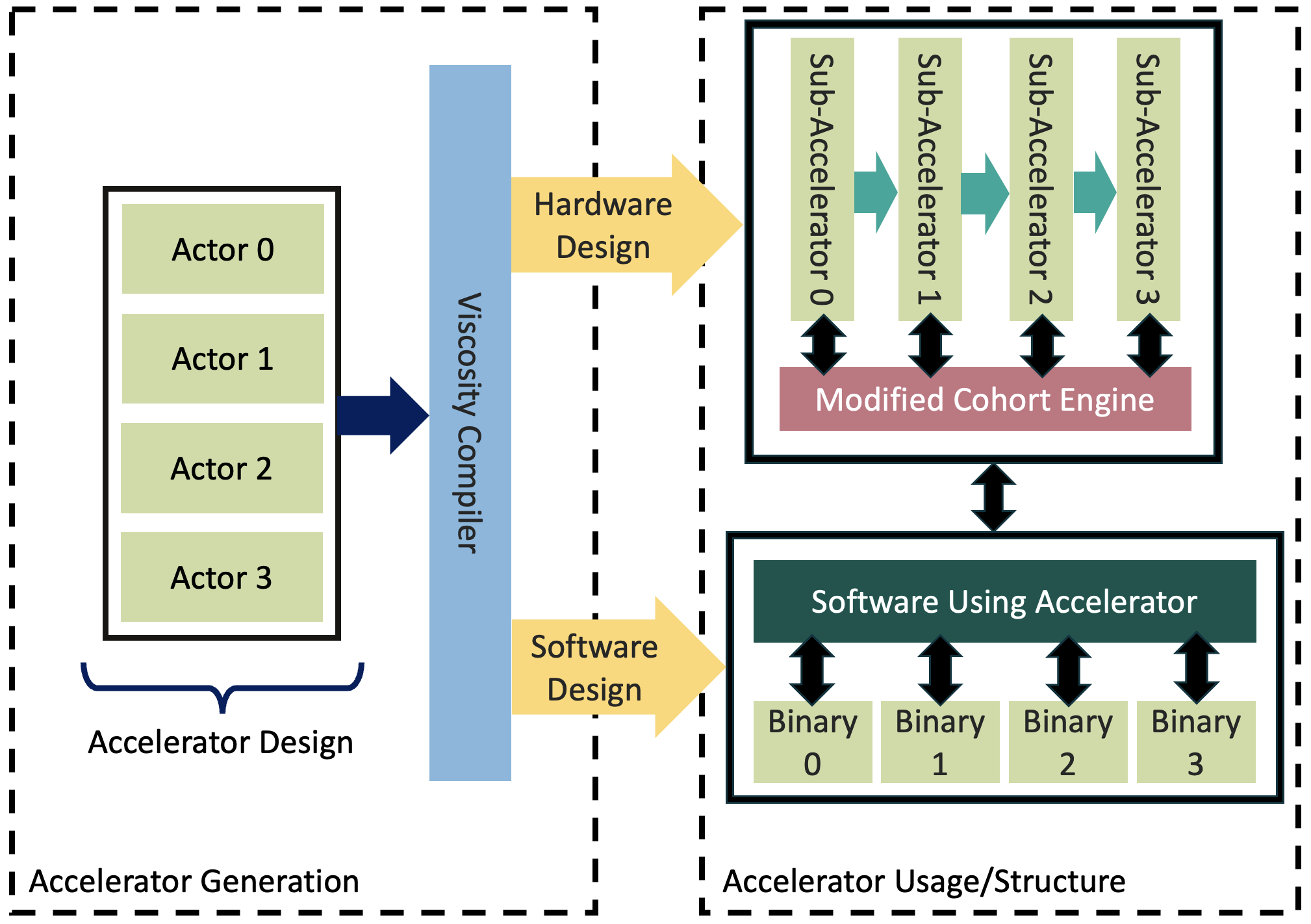}
    \caption{The structure of the proposed accelerator generation and architecture}
    \label{fig:front-page-figure}
\end{figure}
Since the required life-cycle of hardware is increasing, but the practical lifetime of hardware is staying the same or decreasing~\cite{uptime_institute}, novel architectural methods are needed to extend the life of processors. Given the heterogeneity of modern processors, these methods should be targeted to specific parts of the processor to ensure optimal functionality chip-wide. 

We choose to focus on hardware accelerators, which have been the key to improvements in performance for various domain specific applications~\cite{10.1145/3569052.3580028, 10.1145/2554688.2554787, 10.1145/3476229}. Moreover, they have become ubiquitous in modern data center processors~\cite{081eb318add24947a85892d02b1f1cc0, doi:10.1126/science.aam9744}. Often in data center contexts, if a hardware accelerator breaks, the whole processor would need to be replaced. In order to protect hardware accelerators from non-transient device errors, fault tolerance methods achieve tolerance by adding hardware redundancies to the accelerators. However, the area overhead of current methods are simply too expensive, especially as accelerators grow in size and complexity. 

Without fault tolerance, when a fault does occur inside an accelerator, in the worst case, an accelerator must be replaced by a fully software implementation of the same algorithm. In contrast to this method, we propose Oobleck, an architectural design principle leveraging modular design to increase the fault tolerance of accelerators.

For an accelerator composed of a number of sub-accelerators, only the failing sub-component may need to be replaced with a software implementation, but this means that a full software re-implementation of each sub-accelerator would still be needed to handle all failure cases. Generating such an implementation from an RTL implementation of an accelerator would be a challenge and likely limit software performance to that of RTL simulation. 

To mitigate this, we created \textbf{Viscosity}, an accelerator design language (ADL) that produces both Verilog HDL and C code. We argue that better ADLs can encourage compositional design and further take advantage of such design practice to generate efficient software fallbacks for failing sub-accelerators.
 
Our contributions in this paper are fourfold:
\begin{itemize}
\item {\textbf{Models showing how modular fault tolerance can decrease the cost of data center operation by reducing chip replacements.}}

\item{\textbf{The Oobleck Methodology} leveraging modular acceleration within an SoC to gain robust fault tolerance.}

\item {\textbf{Viscosity, a novel hardware-software description language for designing modular accelerators.} Viscosity lowers its custom ADL to both Verilog and C.} 

\item {\textbf{Evaluations of the fault tolerance architecture, including specific examples and generalized benchmarks.} We show how the Oobleck methodology can mitigate the effects of faults on FFT, AES, and DCT accelerators. This method can achieve $1.7\times-5.16\times$ speedup over software implementation after the occurrence of a fault. We show that hot-spare FPGAs, FPGAs embedded into a chip, have the potential to achieve up to 80\% of the original accelerator speed.} 
\end{itemize}

\section{Motivation}
\label{sec:motivation}
When a non-transient error occurs in a data-center accelerator, the data center can either replace the chip that the accelerator is on or eat the performance cost of losing the accelerator. Neither the financial cost nor the hefty increase in operation latency is a good outcome. However, by mitigating the latency increase caused by a fault, we can keep most of the performance of the accelerator and maintain most of the aggregate throughput. One way to achieve this is to make the accelerator fault in stages. The first fault causes some small degradation in performance and each subsequent fault makes the degradation worse until the accelerator no longer functions and the chip must be replaced. We call these "variable fault accelerators" (VFA) which generalise the Oobleck architecture we propose. These are in contrast to accelerators that simply fail upon the first fault and need to be replaced, as is seen today. We call these "single fault accelerators" (SFA).

\begin{figure}
    \subfigure[How the number of replaced chips change dependent on the rate of failure of components. (Lower is better)]{
    \centering
        \includegraphics[width=0.485\textwidth]   {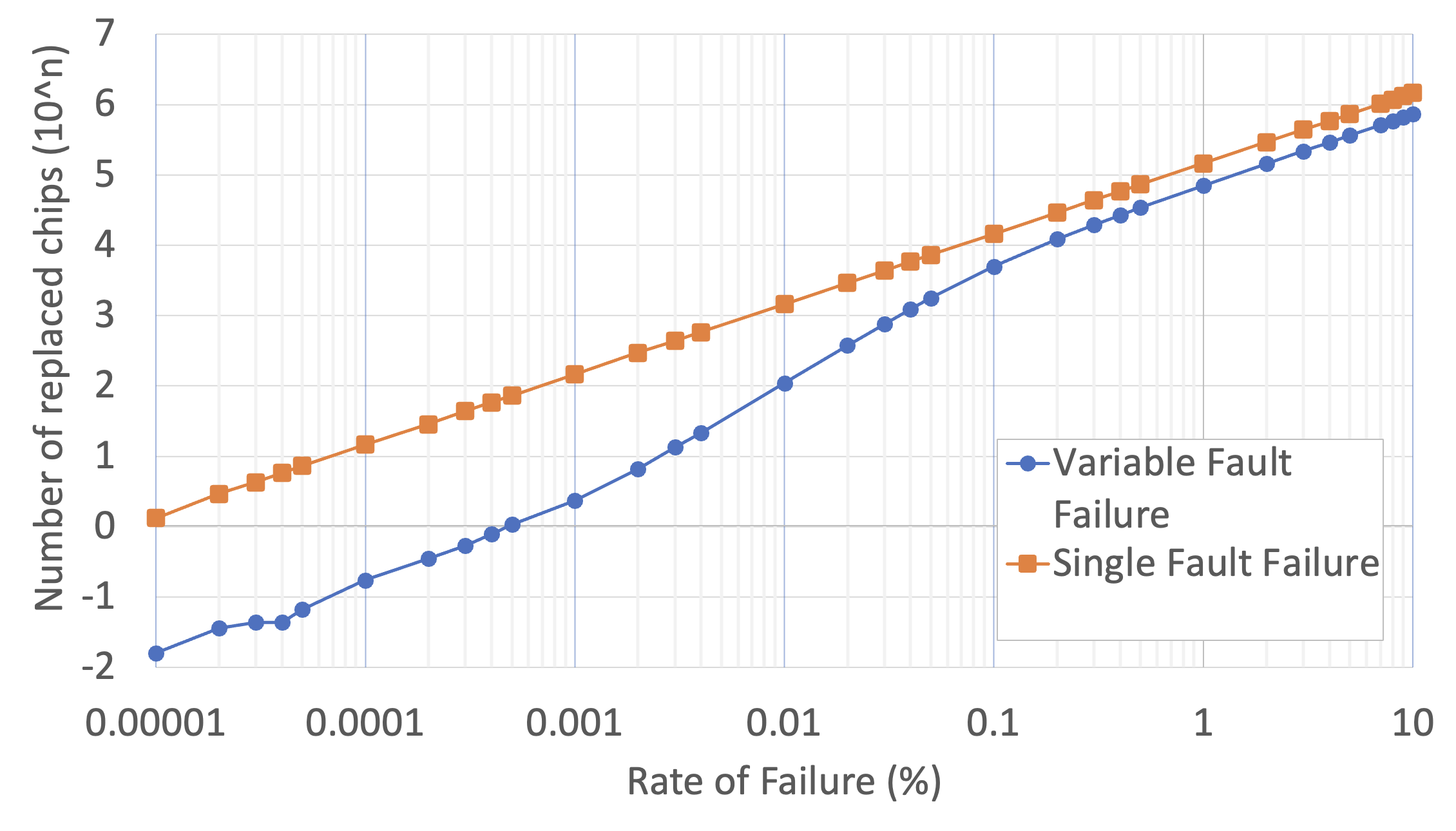}
    }
    
    \subfigure[How the aggregate throughput of the data center changes dependent on the rate of failure of components. (Higher is better)]{
        \centering
        \includegraphics[width=0.475\textwidth]{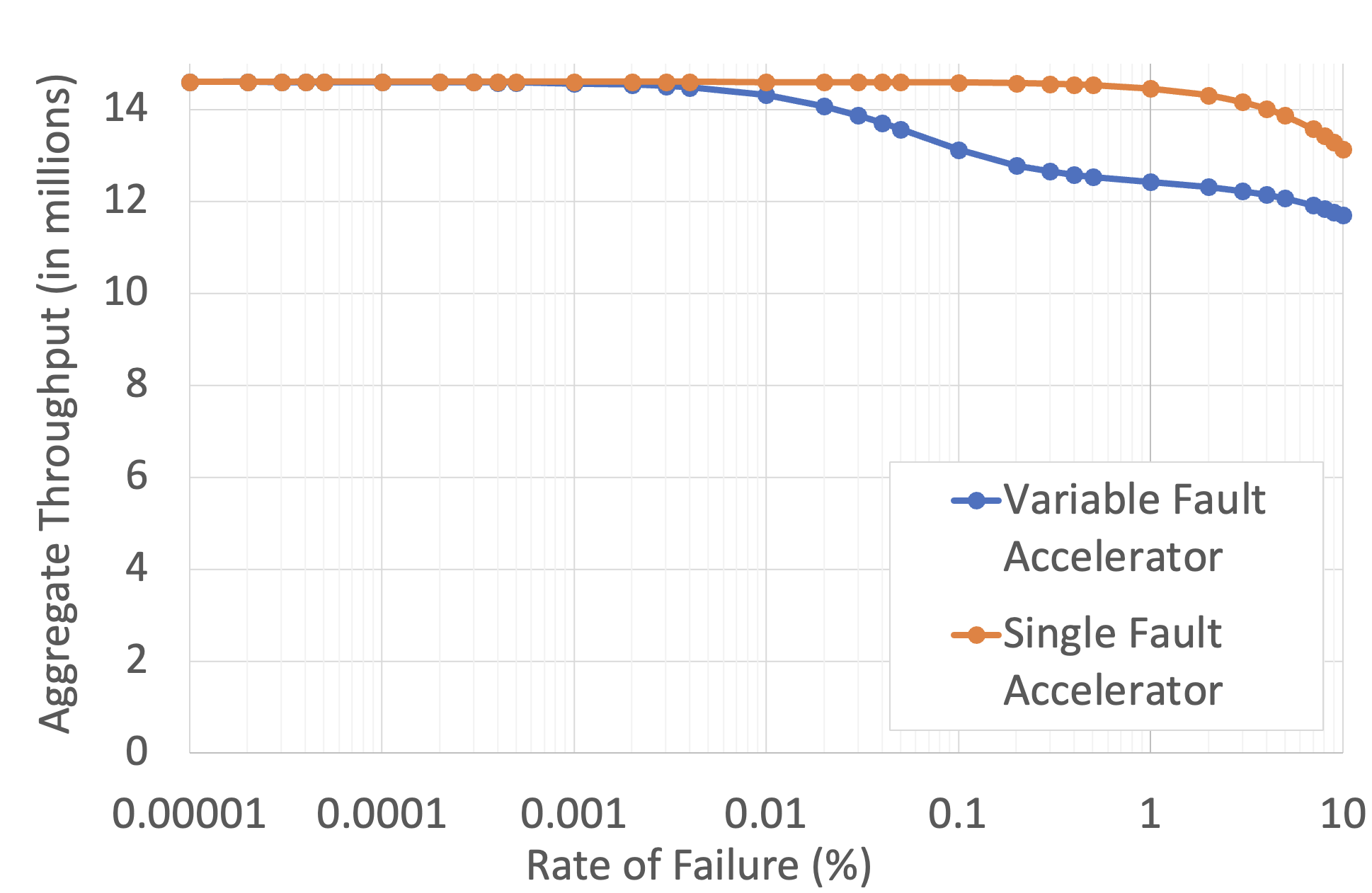}
    }
    \caption{Data center with 10,000 processor data center running over 1460 ticks (4 years at one tick per day). VFAs are assumed to fail after three faults.
    }
    \label{tab:model_outcome}
\end{figure}

\textit{Fixed-Time Data Center Modelling.} To assess the effect of fault tolerance on data centers and to motivate these VFAs, we model data centers employing these two types of accelerators. We fix the number of chips in the data center while measuring the number of processors replaced during the simulation and the aggregate throughput of the data center.

The results of the model are primarily affected by two parameters: fault likelihood per tick and the number of chips in the data center. We model data centers with 10,000 chips. The primary parameter we vary is the fault likelihood per tick between 10\% per tick and 0.00001\% per tick.

Fig. 2(a) shows how the number of processors which are replaced vary over different fault likelihoods. The number of replaced processors for VFAs is strictly lower than SFAs. As the likelihood of faults approaches zero, the difference between the two categories increases. Note that until the failure rate goes below 0.01\%, the difference in costs is small, but once failures drop below that threshold, the difference significantly increases. Thus by using VFAs, data centers would reduce the number of processors replaced to less than one on average, even at fault likelihoods where using SFAs would result in over 50 replacements.

Fig. 2(b) shows that as the likelihood of fault per tick approaches 0, the aggregate throughput of the VFAs approaches the maximum possible aggregate throughput. The difference in aggregate throughput is extremely small when the likelihood of a fault is under 0.01\%.
The model shows that VFAs reduce the number of replaced processors in data centers while not meaningfully reducing throughput.

\begin{figure*}
    \centering
    \includegraphics[width=0.9\textwidth]{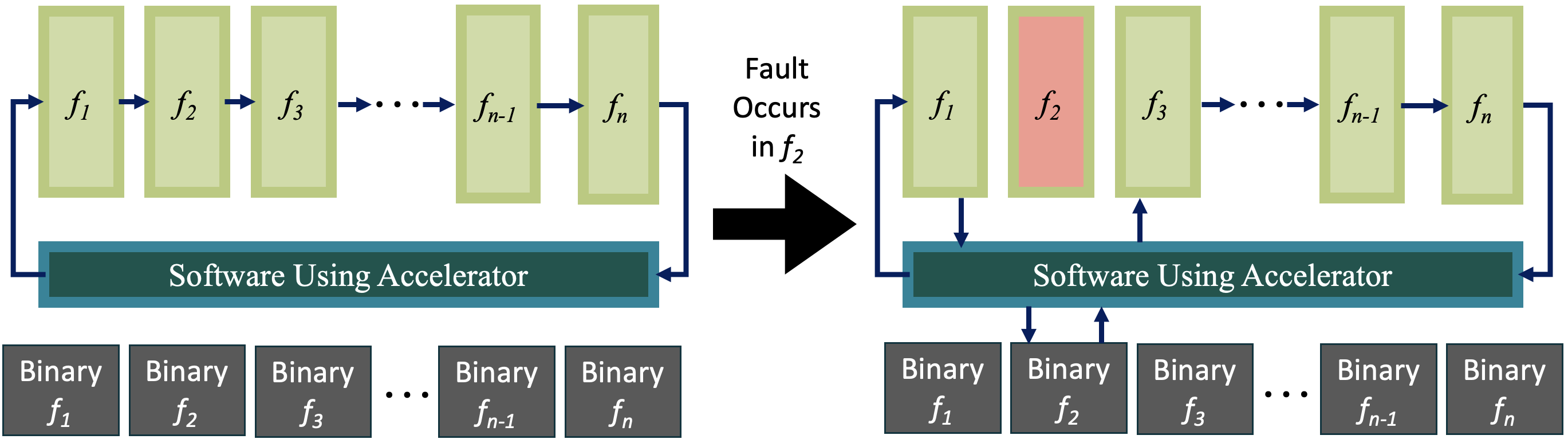}
    \caption{The structure of a modular fault tolerant accelerator before and after a fault occurs.}
    \label{fig:fault_diag}
\end{figure*}

\textit{Fixed Throughput Data Center.} 
The above model uses a fixed number of processors, but we also consider a fixed throughput model where we fix the aggregate throughput of the data center and examine how many chips have to be replaced to maintain that throughput. In a VFA data center, we keep partially faulted accelerators and add new processors to make up the difference in throughput. We have to buy fewer accelerators since we keep the performance of the partially functional VFAs instead of getting rid of the accelerators entirely.

We would have to buy 50\% fewer accelerators if each accelerator only lost half of its performance when a fault occurred. With a one third performance loss, we only have to replace a third of the accelerators that we would otherwise. In general, the number of accelerators that would have to be bought linearly decreases as the efficacy of variable fault processors increases. When evaluating our VFA implementation, we tailor an estimate of cost savings to the performance of the Oobleck architecture.


\section{Our Approach}
\subsection{The Oobleck Methodology}

Currently, for some on-chip accelerator implementing a function \(f\), we suppose a single set of interfaces for input and output. If a non-transient fault occurs in some arbitrary location inside the accelerator, the accelerator will produce the incorrect output despite the fact that the fault itself may make up an extremely small portion of the total accelerator design. Ideally, there should be a way to utilise the unbroken logic within the accelerator, processing only the broken step using some other method of computation.

In order to utilise the logic before and after the fault, the fault logic needs to be isolated. Naturally, this presents an issue because the fault could be at any location and cannot be predicted before taping out the chip. To accommodate the uncertainty in the location of the fault, the accelerator can be split into sub-accelerators. These sub-accelerators add a second set of interfaces. The new interface binds sub-accelerators to their neighbors via latency insensitive interfaces. The old interface to software remains, allowing sub-accelerators to act independently using interfaces which communicate with the kernel driver. In the absence of a fault, the sub-accelerators cohesively communicate with their latency-insensitive interfaces with each other, and only the first and last sub-accelerator interact with software. In the presence of a fault, the cohesiveness can be broken up, allowing isolation of the faulty logic.

Ideally, the hardware design process (and the language in use) would enforce a similar pattern of design which makes fault tolerance most straightforward. We designed Viscosity with the actor model of computation in mind to enforce this modular paradigm. In the actor model, isolated computational units, called "actors," communicate with each other via asynchronous "mailboxes" mirroring the sub-accelerator paradigm that we require.

To build a fault tolerant accelerator using Viscosity which executes a function $f$, create sub-accelerators capturing functions $f_1, f_2, \ldots, f_n$ such that $f_n \circ \ldots \circ f_2 \circ f_1 \ \equiv f$.
Each sub-accelerator has two sets of interfaces, one set of interfaces with the software thread and one set of interfaces with the previous and following sub-accelerators.

Fig.~\ref{fig:fault_diag} shows an example of the system where a fault occurs in the second stage of a modular accelerator. Instead of routing the data through the faulty logic, data is routed from the accelerator through a software binary mirroring the second stage of the accelerator and then back to the next stage of accelerator. The rest of the accelerator can run normally using the interface between sub-accelerators.

Our implementation of the Oobleck methodology runs on a modified version of the Cohort Engine~\cite{10.1145/3582016.3582059}. Cohort eases the complexity of adding of new accelerators while keeping system-level guarantees by providing FIFO queue endpoints for communication between software threads and accelerators built on top of cache-coherent memory queues. Whereas the Cohort Engine supports a single queue per tile, our modified version supports multiple queues interfacing with multiple sub-accelerators. Additionally, we introduce latency insensitive queue-bypassing to enable sub-accelerators to communicate with each other directly. 
The Cohort consumer and producer queues interact with the software thread while the queue-bypass interfaces communicate directly with the previous and following sub-accelerators.

\lstset{frame=tb,
  aboveskip=5mm,
  belowskip=5mm,
  showstringspaces=false,
  columns=flexible,
  basicstyle={\small\ttfamily},
  otherkeywords={@ true, false},
  emph={let, module, int, bool, uint64_t, uint32_t, struct, typedef, pub, derive},
  breaklines=true,
  breakatwhitespace=true,
  tabsize=3,
}
\begin{figure*}[t]
    \label{code.1}
    \noindent\begin{minipage}{\linewidth}\begin{lstlisting}[language=Mathematica,frame=tb]
module [checksum_reg : int = 0] pipelined_checksum (input: int) -> (out : int) {
    let x : int = (input & 0x5555555555555555) + ((input >> 1) & 0x5555555555555555);
    x = (x & 0x3333333333333333) + ((x >> 2) & 0x3333333333333333);
    @checksum_reg = (x & 0x0f0f0f0f0f0f0f0f) + ((x >> 4) & 0x0f0f0f0f0f0f0f0f);
    let y = (checksum_reg & 0x00ff00ff00ff00ff) + ((checksum_reg >> 8) & 0x00ff00ff00ff00ff);
    y = (y & 0x0000ffff0000ffff) + ((y >> 16) & 0x0000ffff0000ffff);
    y = (y & 0x00000000ffffffff) + ((y >> 32) & 0x00000000ffffffff);
    out = y;
} <(y != 0); true>

    \end{lstlisting}\end{minipage}
    \caption{Viscosity code for a Checksum module. The module shows the basic functionality of the Viscosity, including operations, state variables (registers), and the valid/ready outputs.}
    \label{fig:visc1}
\end{figure*}

When a fault occurs, the change in accelerator latency is affected by three primary factors: the number of cycles to move data between the software thread and the accelerator, the ratio of cycles to run the function in hardware versus software, and the number of stages comprising the accelerator. The added latency is roughly the difference in latency between the software and hardware versions of the sub-accelerator added to the transmission latency (the latency of moving data between the software thread and the hardware accelerator).


Our modified Cohort Engine allows sub-accelerator configuration from software. Each accelerator accepts a two bit configuration signal where the higher bit represents whether the accelerator should wait for data from the consumer queue and the lower bit represents whether the accelerator should push the data to the producer queue (otherwise sub-accelerators are directly connected). 

\textbf{Oobleck does not dictate a particular method of fault detection}; any method which can communicate state with the accelerator, either a software or hardware mechanism, can be used.

\subsection{Viscosity}

The Oobleck methodology requires both a software and hardware version of each sub-accelerator. There are three reasons we would want to generate both the hardware and software from a single description: Firstly, it allows the hardware engineer to only describe the operation once. Secondly, it ensures that the software and hardware versions of the operation are logically equivalent, especially important given the complexity of subdividing hardware modules. Lastly, it lets the language enforce the sub-accelerator modularity convention.

Generating both a hardware and software description from a single description could be done with hardware simulation tools such as Verilator, VCS, or Vivado. However, the high overhead of hardware simulators makes them slower than a software optimized version of the same operation. Additionally, HLS solutions do not allow the granularity of splitting needed to keep the modules evenly timed.

The best alternative to custom written software would be generated software that could fit inline with the user space code using the accelerator. We created our own language, \textbf{Viscosity}, which compiles to C and to Verilog via the Shakeflow HDL~\cite{10.1145/3575693.3575701}. Shakeflow is a Rust-based DSL which improves on previous functional HDLs by adding latency-insensitive interface combinators which model the accelerators we target well. 


\textit{Usage}. Each module generated using Viscosity supports both combinational and sequential logic.
To build combinational logic, declare a module as shown below, with the list of input signals followed by the list of output signals.

\noindent\begin{minipage}{\linewidth}
\begin{lstlisting}
module [] <operation_name> (<input_name> : <type>, ...) -> (<output_name> : <type>, ...) {
    let int_variable : int = ...; 
    let bool_variable : bool = ...;
    let arr_variable : [<number of elements>] = ...;
} <true; true>

\end{lstlisting}
\end{minipage}

Each module allows boolean expressions to set the conditions of the valid-ready interface. To set these, Viscosity has a post-function script, which has access to the entire scope of the function body. Fig.~\ref{fig:visc1} shows the pipelined\_checksum module with two expressions as \lstinline{<(y != 0); true>}. The valid parameter is set to \lstinline{(y != 0)}, meaning that the module will emit a valid signal if the variable \lstinline{y} inside the module does not evaluate to 0 at the end of the cycle.



In order to build a sequential logic module, we introduce state variables. A state variable is set at the beginning of a cycle and is updated before the end of that cycle to be used in the following cycle. You can reference the previous state value by referencing the variable directly and you can assign the next state value by using the \lstinline{@} operator as shown below.

\noindent\begin{minipage}{\linewidth}\begin{lstlisting}
module [<reg_name> : <type> = <inital value>, ...]
    <operation_name> (input_name : <type>, ...) -> (output_name : <type>, ...) {
    let old_register_value : <type> = <reg_name>;
    @<reg_name> = <expression>;
} <true; true>

\end{lstlisting}\end{minipage}

\subsubsection{Compilation to Shakeflow HDL}

\textit{Input, Output, and State Interfaces}. Shakeflow represents the inputs and outputs as structs derived from the signal type. A set of inputs, outputs, and state variables will be translated into 
signal interfaces.

\noindent\begin{minipage}{\linewidth}\begin{lstlisting}
#[derive(Debug, Clone, Signal)]
pub struct I_<operation_name>{
	<input_0>: <type_0>, ..., <input_n>: <type_n>
}
#[derive(Debug, Clone, Signal)]
pub struct O_<operation_name>{
	<output_0>: <type_0>, ..., <output_n>: <type_n>
}

// If there is sequential logic}
#[derive(Debug, Clone, Signal)]
pub struct State_<operation_name>{
	<reg_0>: <type_0>, ..., <reg_n>: <type_n>
}
\end{lstlisting}\end{minipage}

\subsubsection{Compilation to C}


\textit{Input, Output, and State Interfaces}. For C compilation, we only need a struct for the output. The output variables are compiled to the following struct and type. 


\noindent\begin{minipage}{\linewidth}\begin{lstlisting}
struct _<operation_name>_output {
    uint64_t integer_output;
    bool boolean_output;
    uint64_t array_output[ARRAY_LENGTH];
};
typedef struct _<operation_name>_output <operation_name>_output;

<operation_name>_output <operation_name> (<type_0> <input_name_0>, ...){...} 
\end{lstlisting}\end{minipage}
\begin{figure*}
    \centering
    \includegraphics[width=\textwidth]{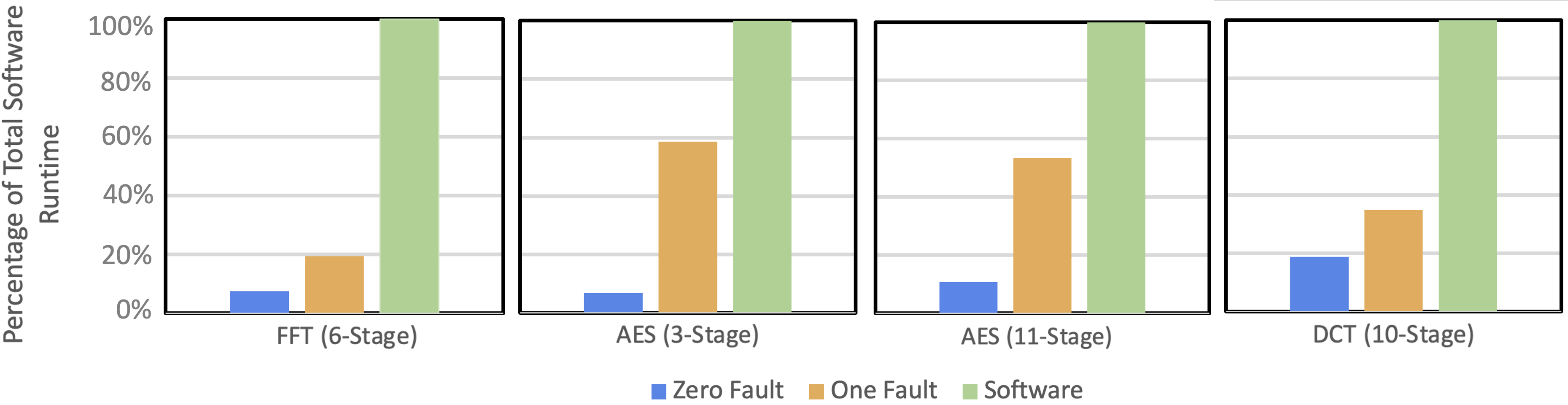}
    \caption{Execution time of various case studies as a percentage of software execution time}
    \label{fig:fftaesdct}
\end{figure*}
When there are no state variables and the logic is combinational, the body of the function executes the operations on the inputs, putting the outputs into the declared struct and returning it.
When the logic is sequential, a wrapper for the function is created which runs in a while loop which iterates over a set of inputs and state variables for each cycle, outputting once the valid signal evaluates to true.

\section{Evaluation Methodology}

\renewcommand{\arraystretch}{1.3}

We implemented various accelerators using the Oobleck methodology with a modified Cohort Engine on the OpenPiton+Ariane RISC-V Research Platform~\cite{10.1145/3582016.3582059}. We then booted Linux (v6.2, built via Buildroot) on a Digilent Genesys2 FPGA (Kintex-7 XC7K32T5-2FFG900C) running at a clock speed of 67 MHz. All of the evaluation methods use this configuration. We run four different evaluations to test the efficacy of the fault tolerance mechanisms:

\textit{Case Studies.} Our first evaluation focuses on evaluating how the modular accelerators function on specific accelerator architectures: Fast Fourier Transform (FFT), Advanced Encryption Standard (AES), and Discrete Cosine Transform (DCT) accelerators. All three are written in Viscosity and compiled to Verilog and C. Each accelerator is split into stages where each stage can take an arbitrary number of cycles. Our baseline of comparison is the software version of these written in C. We compare the performance of these accelerators when we simulate a fault to a software version of the same operation.

\textit{Pass-Through Accelerator.} For our second evaluation, we use an accelerator which emulates the latency of an operation while allowing data to simply pass through. We specify how long we want to spend in the accelerator hardware and the corresponding software fallback. For each configuration we compare the performance against the cumulative runtime of the software.

\textit{Pass-Through Accelerator with Multiple Faults.} For our third evaluation, we use the same hardware setup but consider the performance under two faults, testing with a subset of the previously considered configurations. 

\textit{Pass-Through Accelerator with Hot-Spare FPGA Fallback.} Our last evaluation tests how the performance of the pass-through accelerator changes when we use a hot-spare FPGA as a fallback instead of software. In order to evaluate the performance we add a new parameter, FPGA speedup over software.

\begin{table}
    \centering
        \caption{Sizes of the Tested Accelerators}
    \begin{tabular}{|c|c|c|}
        \hline 
        Accelerator & LUTs & Slice Registers \\ 
        \hline
        FFT (6-Stage) & 36,959 & 24,973\\
        AES (3-Stage) & 11,725 & 2,314 \\
        AES (11-Stage) & 23,183 & 15,210 \\ 
        DCT (10-Stage) &  24,808 & 20,723 \\ 
        \hline
    \end{tabular}
    \label{tab:my_label}
\end{table}
\section{Results}

\subsection{Fast Fourier Transform Accelerator}

We implemented a Fast Fourier Transform accelerator with Viscosity following the Oobleck methodology. The FFT uses a butterfly design where each stage of butterflies is one stage of the resulting accelerator. 

The execution time for the FFT accelerator with zero faults and with one fault are shown on the left of Fig.~\ref{fig:fftaesdct} as a percentage of the software runtime of the same function.
When there is no fault present, the FFT Accelerator performs with only 7.4\% of the cycles when compared to the software implementation, a speedup of $13.5\times$. When there is a single fault present, the FFT Accelerator runs in approximately 19.3\% of the cycles of its purely software counterpart, a speedup of $5.181\times$. 

\subsection{Advanced Encryption Standard Accelerator}

We designed an AES accelerator with two different configurations: an 11-stage AES accelerator and a 3-stage AES accelerator. In the 3-stage configuration we organised the operations such that the key expansion and first two rounds are run in the first stage and four rounds are run in each of the next two stages. For the 3-stage and 11-stage accelerator, each stages software fallback takes $\approx$17,788 cycles and $\approx$5,000 cycles respectively. The small size of the accelerator lowers the potential benefit of using the Oobleck structure.   

We found that the number of stages has generally no effect on the performance of the AES accelerator. This is due to the low number of cycles that the AES operation took in software when compared to the speed of the Cohort Engine. 
When a fault does occur in the hardware and the software fallback takes over, the efficiency of the system drops more with the accelerator taking 58\% of software's execution time under a fault.

\subsection{2-D Discrete Cosine Transform Accelerator}

Lastly, we analysed the performance of computing a 2-D Discrete Cosine Transform under fault. This DCT accelerator utilises a modified 10-stage butterfly design with several key optimisations in its internal structure specifically for the JPEG compression algorithm. Compared to the other accelerators we evaluated, the DCT accelerator has the lowest software to hardware cycle ratio, which is due to the design already leveraging the fastest known DCT algorithm. 

Even with this performance ceiling, the accelerator with no faults still runs in approximately 18.9\% of the total software cycles for a speedup of $5.3\times$. With one faulty stage, the DCT accelerator records a speedup of $2.87\times$, an encouraging performance as the software implementation is already heavily optimised.
\begin{figure*} 
    \centering
    \includegraphics[width=0.9\textwidth]{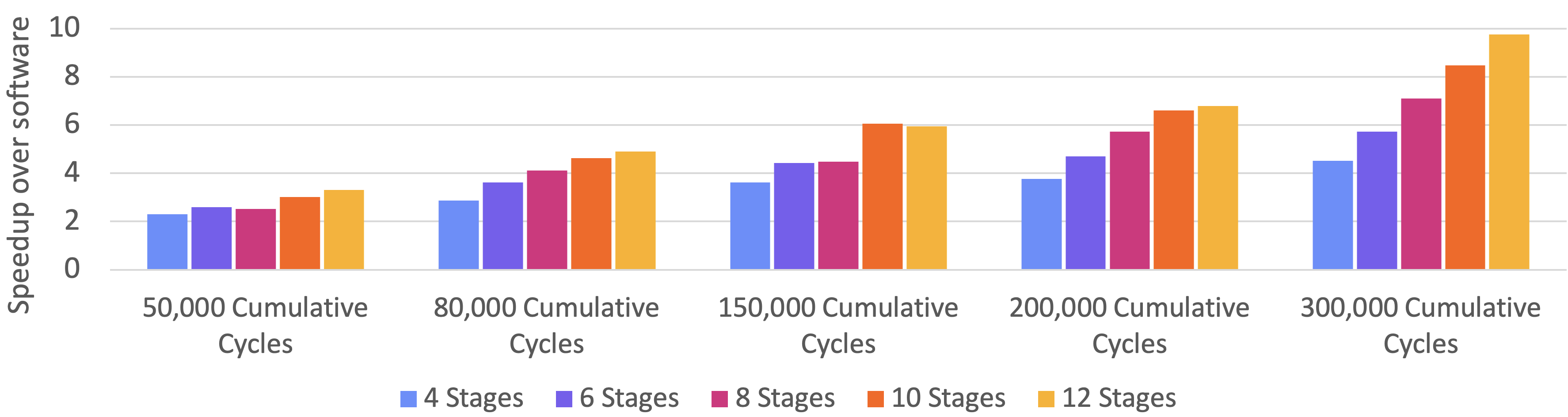}
    \caption{Speedup of a pass-through with one faulty stage over a purely software implementation. Each hardware stage is assumed to take 100 cycles.}
    \label{fig:perfprof}
\end{figure*}

\subsection{Performance Profiling}

From the case studies, we identified three primary parameters which affected the efficacy of the accelerators under fault: the number of cumulative software cycles (how much time does the entire operation take to run in software), the speedup of the accelerator over software, and the number of stages in the accelerator. In order to sweep the design space between these three variables, we use a pass-through accelerator.

Fig.~\ref{fig:perfprof} shows how the number of stages and size of operation affects performance when we fix the size of the operation. We assume that the accelerator has a speedup of $100\times$ over software. That means for the 300,000 cumulative cycle operation, the accelerator latency is 3,000 cycles not including transmission latency between the software and accelerator.

As either the number of stages or the size of the operation increases, so does the speedup over software when a fault has occurred (as less of the operation must be done in software when the fault occurs). Moreover, the sensitivity of performance to the number of stages is dependent on the size of the operation. For example, the change in speedup for the smallest operation when tripling the stages is 1 (from $2.3\times$ to $3.3\times$) whereas for the largest application it is 5.2 (from $4.5\times$ to $9.7\times$).




\subsection{Performance Under Multiple Faults}
\begin{figure} 
    \includegraphics[width=0.43\textwidth]{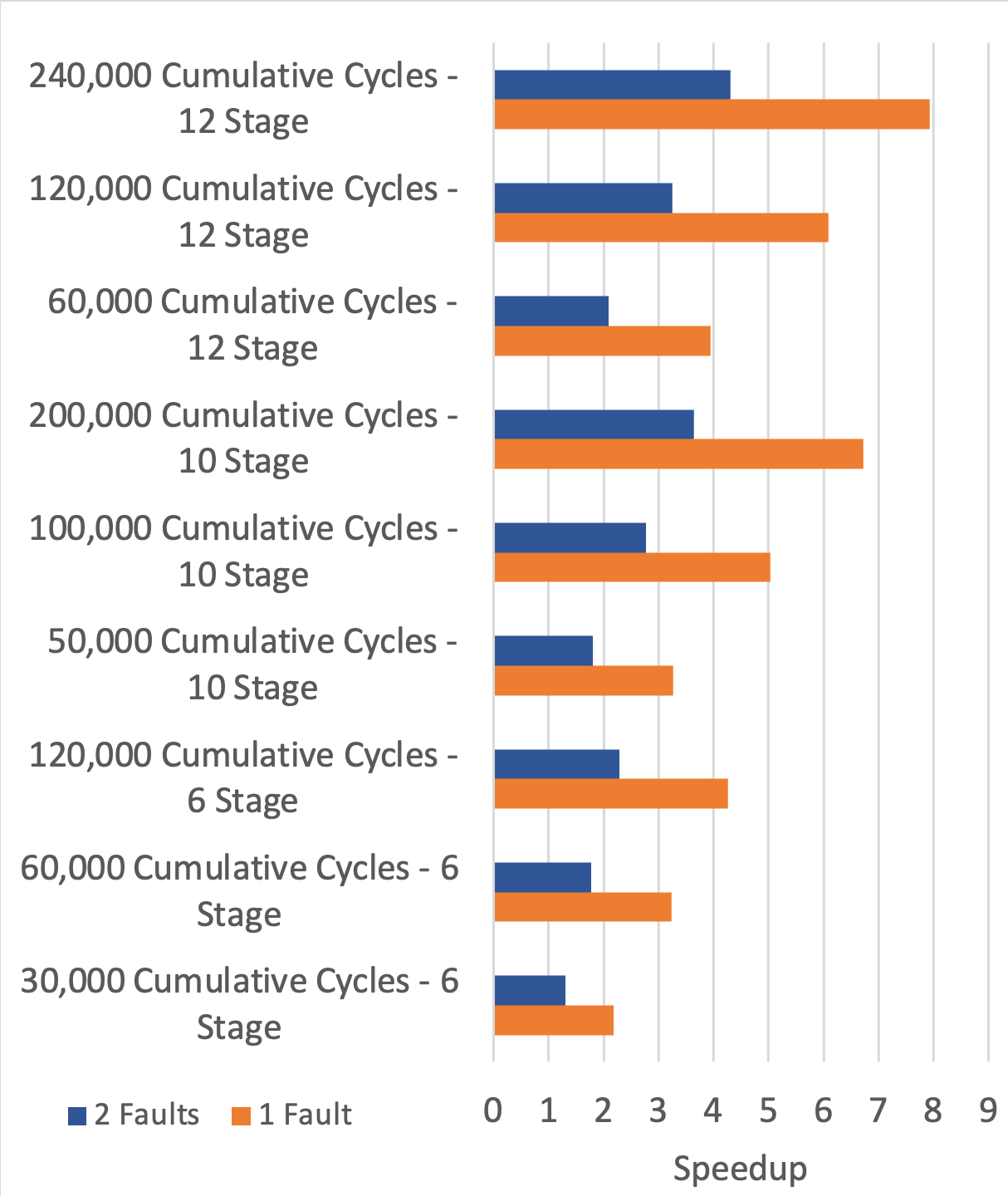}
    \caption{Speedup of pass-through accelerator with two faults over purely software implementation.}
    \label{fig:multifault}
\end{figure}
We additionally profile the system under multiple faults. We run a modified version of the above experiment, mirroring all the parameters and simply changing the number of stages in which a fault occurs.

Fig.~\ref{fig:multifault} shows that when there are two faults in the accelerator, it is still faster than the purely software version. However, when the number of cumulative cycles for the software implementation is very low, the two fault accelerator does not maintain significant speedup over the software version. For an accelerator running only 30,000 cumulative cycles split into 6 stages, the two fault speedup is only $1.3\times$ as opposed to the one fault speedup of $2.17\times$. As the number of cumulative cycles for the software implementation increases, the speedup when there are two faults in the accelerator increases, maintaining about half of the speedup of the accelerator with only one fault. For the 240,000 cumulative cycle accelerator split over 12 stages, the accelerator maintains a speedup of $4.30\times$. For the 200,000 cumulative cycle accelerator split over 10 stages, the speedup under two faults is $3.65\times$.

With the exceedingly rare case of more than two faults occurring in different sub-accelerators, there may be situations where there is no speedup to bypass further stages. For example, the 30,000 cumulative cycle accelerator split into 6 stages would likely be slower in the presence of a third fault than a software implementation of the same operation due to transmission latency. The 240,000 cumulative cycle accelerator split over 12 stages could have up to 8 faults and still perform better than the software implementation.




\subsection{Opportunities for Further Acceleration Using Reconfigurable Fabrics}

Since a Verilog description is generated by Viscosity, an FPGA bitstream for the specific sub-accelerators could also be generated. Then, in the event of a hardware failure, the hardware could fall back to a reconfigurable fabric configured to that bitstream. This would outpace the software fallbacks if the reconfigurable fabric is already configured to that sub-accelerator. The number of stages which could use the reconfigurable fabric as a fallback would depend on the size of the fabric.


We estimate the speedup of using a hot-spare FPGA connected though a Cohort Engine by varying the speedup of the FPGA over software. FPGA speedup over software has been show to range from 35 to over 200 times speedup~\cite{10.1145/968280.968304}. For this estimation, we use an operation which takes 60,000 cumulative cycles in software and is split into 6 stages. The number of cycles for an FPGA is determined by taking the number of software cycles per stage and dividing it by the FPGA speedup factor.

\begin{figure}
    \centering
    \includegraphics[width=0.45\textwidth]{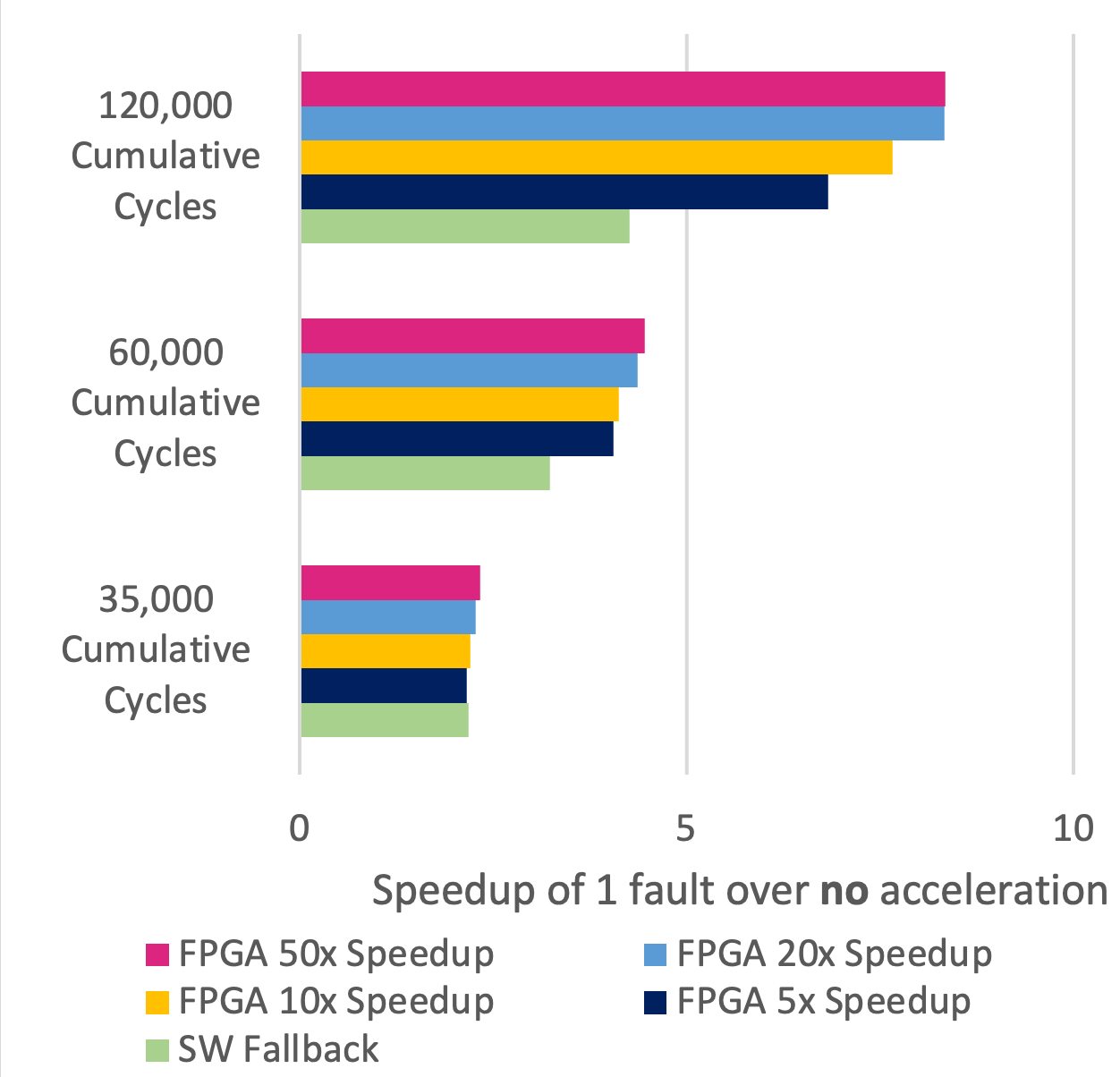}
    \caption{Speedup of FPGA fallback over SW fallback on a 6-Stage modular accelerator when data is routed through software.}
    \label{fig:fpgafallback}
\end{figure}

Fig.~\ref{fig:fpgafallback} shows that the FPGA speedup does not make a significant difference when the operation is small but can lead to significant speedup when the operation is large. This is because (as opposed to software) the movement of data from the software to FPGA is bottlenecking the performance of the FPGA, even when is it has large improvements over the software version. When there is an FPGA fallback, the actual processing takes up a very small percentage of the total processing time. Most of the cycles will be spent on moving data between hardware and software. The performance of the FPGA fallback is largely affected by latency of the Cohort Engine. 

\subsection{Cost Evaluation}

\begin{figure*}[!t]
\centering
\includegraphics[width=\textwidth]{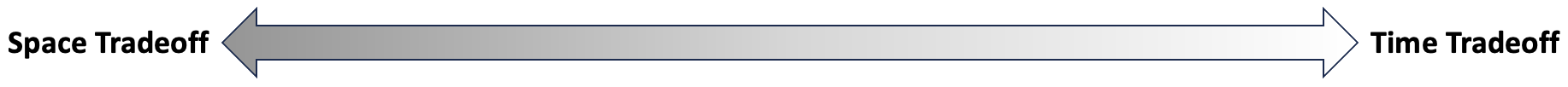}

\subfigure(a){
\centering
\includegraphics[width=0.3\textwidth]{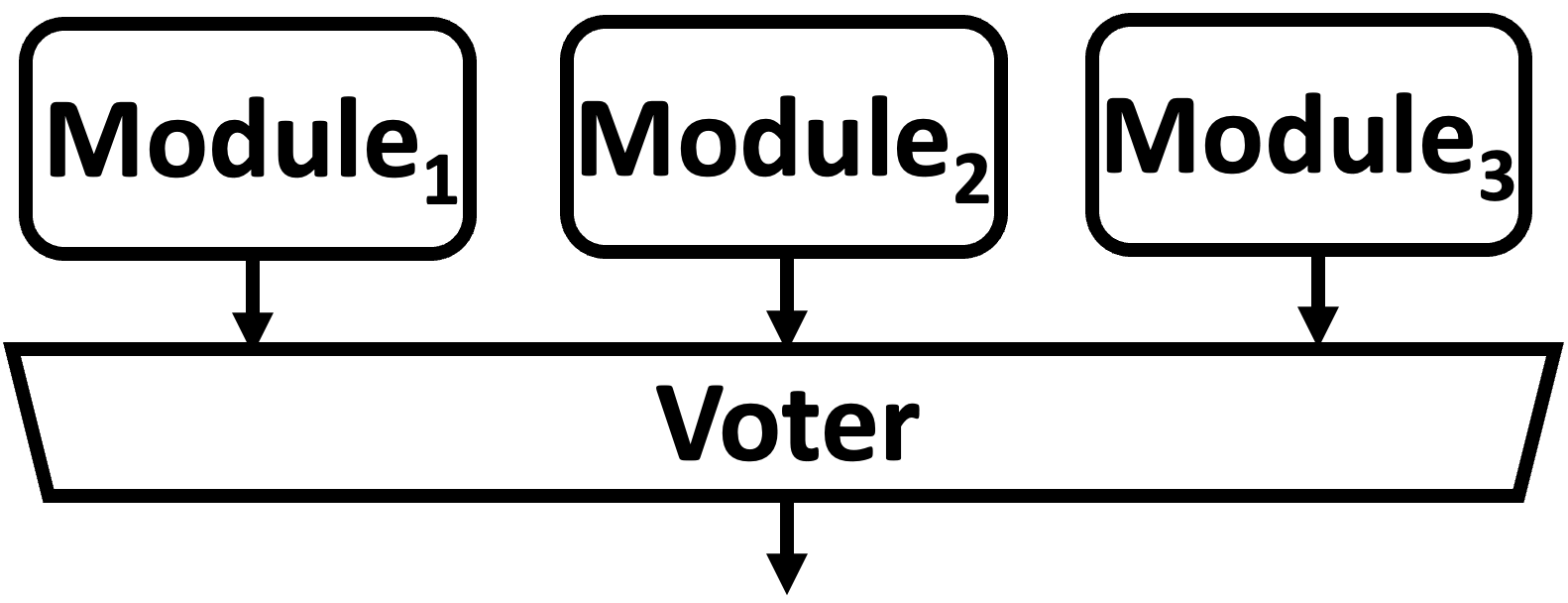}
}\subfigure(b){
\centering
\includegraphics[width=0.3\textwidth]{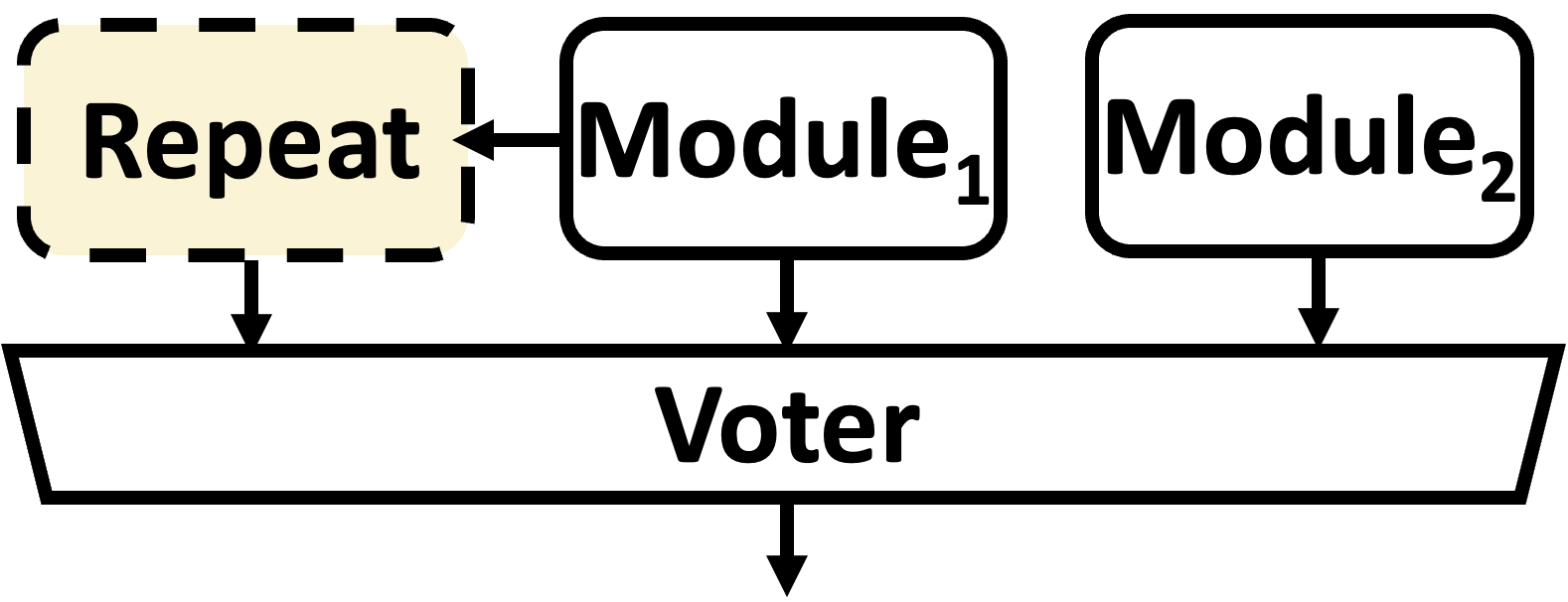}
}\subfigure(c){
\includegraphics[width=0.3\textwidth]{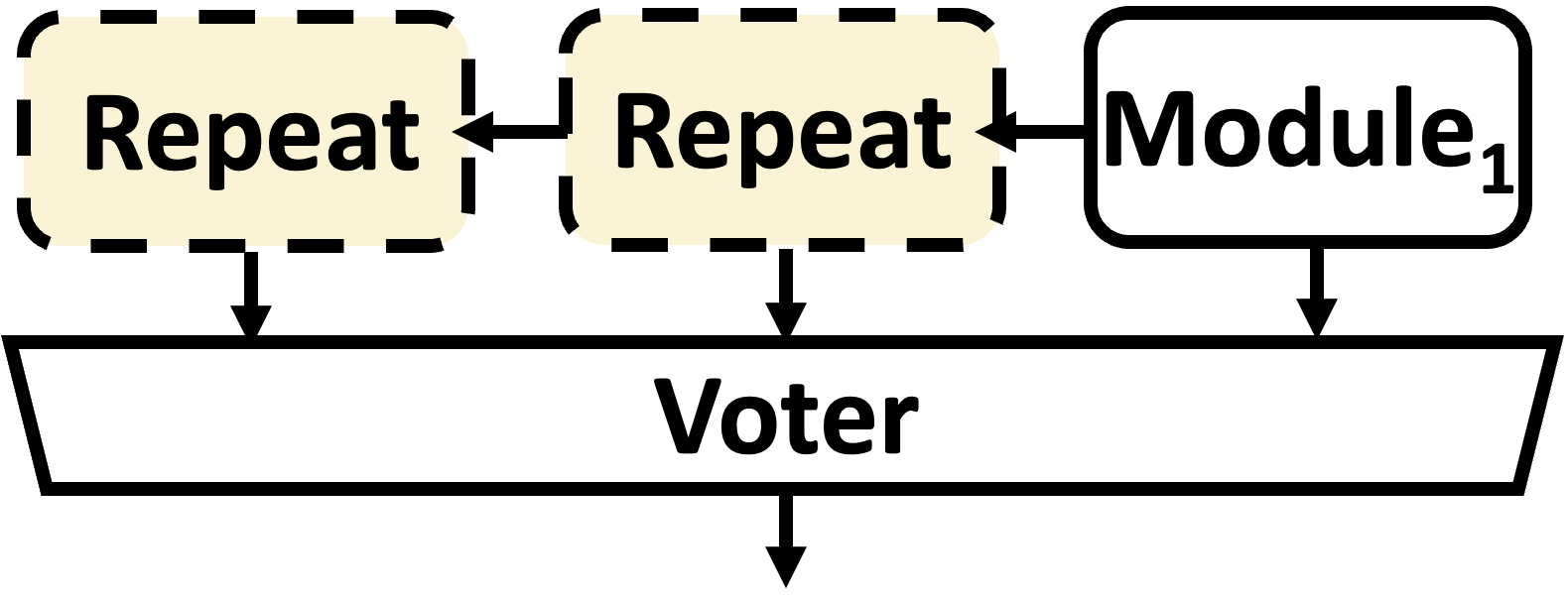}
}

\caption{(a) A Triple Module Redundancy, a type of Space Redundancy (b) A Double Module Redundancy with a re-execution, a type of mixed Space-Time Redundancy (c) A single module ran three times, a type of Time Redundancy}

\end{figure*}

We briefly discussed in Section~\ref{sec:motivation} how the performance of a variable fault accelerator would affect the cost of data center operation assuming fixed throughput operation. We see that with software fallbacks, we can expect speedups of up to eight times from software and a slowdown of two-thirds from the purely hardware accelerated version. 

From our data center models, we expect that modular accelerators can reduce the number of replaced processors by one-third. If we use hot-spare FPGA fallbacks connected directly to accelerators, we could reduce replacements by up to 80\%. As the efficacy of the Oobleck methodology (or other methods of VFAs) increases, the potential cost savings in data centers will continue to increase. Sice the efficacy of our proposal is largely affected by the latency of moving data between the software thread and the hardware accelerator, as those mechanisms improve, so will the savings in replacements.

\section{Related Work}
\subsection{Traditional Fault Tolerance for Accelerators}
Traditional fault tolerance generally falls in three categories:
\begin{enumerate}
    \item \textbf{Space Redundancy} trades off area for reliability. Most popular is Triple Module Redundancy (TMR) (Fig. 3(a))~\cite{taher2019fault}.
    \item \textbf{Time Redundancy} trades off cycles for reliability by re-running and comparing outputs (Fig. 3(c))~\cite{taher2019fault}.
    \item \textbf{Mixed Time-Space Redundancy} uses both hardware and re-execution redundancies (Fig. 3(b))~\cite{1336957}.
\end{enumerate}

Time redundancies only provide benefits when the fault is transient, that is, if the fault only occurs for one execution of the accelerator. Our approach conversely only provides benefit for non-transient faults, meaning it does not necessarily replace time redundancies. However, they could be used in tandem with time redundancies to provide more robust protections from all types of faults. 
Naively, space redundancies simply replicate the accelerator, forwarding the outputs into a voting system as seen in Fig. 3(a). Further research into fault tolerance for accelerators has shown that it is possible to reduce the size of the redundancies~\cite{485368}.

\subsection{Related Work}

Most work on fault tolerance in accelerators has come from exploring hardware generated from HLS tools~\cite{10.1145/3299874.3318020}. Karri and Orailoglu proposed building fault tolerant ASICs using HLS~\cite{229924}. They reduced the area trade-off of TMRs by sharing functional units between modules. Other work has mixed time and space redundancies~\cite{7544380} and introduced various methods to better explore the design space such as genetic algorithms~\cite{4211831, 10.1145/3299874.3318020}. Other works have focused on only building redundancies for the most critical paths~\cite{7544380}.

Other methods vary the area of redundancies. We instead explore the design space between modularity and fault tolerance. Since our methodology uses a modular design, it does not require hardware redundancies.

Other works have considered modular acceleration for non-fault tolerance purposes. Frankenstein Accelerators leverage modular acceleration for cloud services~\cite{Frankenstein}. This focuses on composing sub-components called ``computational kernels'' into larger ``Frankenstein'' accelerators. This structure helped increase the utilisation of resources.

\bibliographystyle{IEEEtran}
\bibliography{mybib}

\end{document}